\documentclass[conference]{IEEEtran}
\IEEEoverridecommandlockouts

\usepackage{cite}
\usepackage{amsmath,amssymb,amsfonts}
\usepackage{algorithmic}
\usepackage{graphicx}
\usepackage{textcomp}
\usepackage{xcolor}
\usepackage{svg}
\usepackage{tipa}
\usepackage{colortbl}
\usepackage{multirow,arydshln}
\usepackage{tablefootnote}
\usepackage{balance}
\usepackage{hyperref}

\def\BibTeX{{\rm B\kern-.05em{\sc i\kern-.025em b}\kern-.08em
    T\kern-.1667em\lower.7ex\hbox{E}\kern-.125emX}}
\begin{document}

\title{\fontsize{19}{24}\selectfont Revise, Reason, and Recognize: LLM-Based Emotion Recognition via Emotion-Specific Prompts and ASR Error Correction}


\makeatletter
\newcommand{\linebreakand}{%
  \end{@IEEEauthorhalign}
  \hfill\mbox{}\par
  \mbox{}\hfill\begin{@IEEEauthorhalign}
}
\makeatother


\author{
\IEEEauthorblockN{Yuanchao Li$^1$, Yuan Gong$^\ast$\thanks{$^\ast$Work done at MIT, now with xAI Corp.}$^2$, Chao-Han Huck Yang$^3$, Peter Bell$^1$, Catherine Lai$^1$}
\IEEEauthorblockA{\textit{$^1$University of Edinburgh, $^2$MIT CSAIL, $^3$NVIDIA Research}\\
yuanchao.li@ed.ac.uk}
}

\maketitle

\begin{abstract}
Annotating and recognizing speech emotion using prompt engineering has recently emerged with the advancement of Large Language Models (LLMs), yet its efficacy and reliability remain questionable. In this paper, we conduct a systematic study on this topic, beginning with the proposal of novel prompts that incorporate emotion-specific knowledge from acoustics, linguistics, and psychology. Subsequently, we examine the effectiveness of LLM-based prompting on Automatic Speech Recognition (ASR) transcription, contrasting it with ground-truth transcription. Furthermore, we propose a \textsc{Revise-Reason-Recognize} prompting pipeline for robust LLM-based emotion recognition from spoken language with ASR errors. Additionally, experiments on context-aware learning, in-context learning, and instruction tuning are performed to examine the usefulness of LLM training schemes in this direction. Finally, we investigate the sensitivity of LLMs to minor prompt variations. Experimental results demonstrate the efficacy of the emotion-specific prompts, ASR error correction, and LLM training schemes for LLM-based emotion recognition. Our study aims to refine the use of LLMs in emotion recognition and related domains.
\end{abstract}

\begin{IEEEkeywords}
Emotion Recognition, LLM, ASR, Acoustics, Linguistics, Psychology
\end{IEEEkeywords}

\section{Introduction}
Recent studies have suggested that Large Language Models (LLMs)  have the ability to reason about emotional content \cite{sap2019social}. This finding has encouraged researchers to further explore the ``emotional intelligence'' (e.g., emotion recognition, interpretation, and understanding) of LLMs. For example, \cite{wang2023emotional} developed a psychometric assessment focusing on emotion understanding to compare the emotional intelligence of LLMs and humans. They found that most LLMs achieved above-average Emotional Quotient (EQ) scores, with GPT-4 surpassing 89\% of human participants with an EQ of 117.

Therefore, the use of LLMs in text-based emotion recognition has emerged as a resource-efficient and effort-saving alternative to human annotators and traditional emotion classifiers for two main reasons: \textbf{\textit{1)}} Emotion recognition requires substantial human effort. Typically, multiple annotators are needed for each sample to reach a majority vote, ensuring accurate assessment. Although platforms like Amazon Mechanical Turk provide a relatively efficient solution, concerns persist regarding privacy leaks, subjective bias, and the reliability of annotations \cite{feng2024foundation}. \textbf{\textit{2)}} Despite the advancements of state-of-the-art deep learning technologies, training an emotion classifier involves multiple steps, including feature extraction and model building, which require careful consideration of which features, models, and algorithms to use. 

To this end, researchers have recently started exploring and utilizing LLMs for emotion annotation and recognition. Their work includes various approaches, such as using multiple-step prompting with LLMs \cite{hama2024emotion}, examining prompt sensitivity \cite{amin2024prompt}, integrating outputs from multiple LLMs \cite{feng2024foundation}, and incorporating acoustic information \cite{santoso2024large}. Despite these efforts, understanding of the usage, efficacy, and reliability of LLM-based approaches remains limited, especially on Automatic Speech Recognition (ASR) transcription. Therefore, building upon existing literature, we conduct various experiments on LLM-based emotion recognition to investigate effective prompting practices. Our major contributions are:

\begin{itemize}
    \item We propose novel prompts for LLM-based emotion recognition, integrating emotion-specific knowledge from acoustics, linguistics, and psychology.
    \item We propose the \textsc{Revise-Reason-Recognize} (R3) prompt for emotion recognition on imperfect text, addressing the limitations of prior methods on ground-truth text.
    \item We experiment on LLM training schemes, evaluating the efficacy of context-aware learning, in-context learning, and instruction tuning for LLM-based emotion recognition, and investigate the sensitivity of LLMs to minor prompt variations.
\end{itemize}

\section{Related Work}
\label{sec:relatedwork}

Interest in LLM-based emotion recognition has surged recently, with the availability of pre-trained models. Studies in this area have explored a variety of approaches. \cite{gong2023lanser} inferred emotion labels using three different prompting approaches: text generation, mask filling, and textual entailment, employing a fine-grained emotion taxonomy. \cite{feng2024foundation} proposed an ensemble approach that integrates outputs from multiple LLMs, leveraging a Mixture of Experts (MoE) reasoning model. They trained emotion classifiers using MoE-generated emotion labels from both ground-truth and ASR transcriptions, and tested these classifiers on ground-truth labels, demonstrating comparable performance in emotion classification. \cite{hama2024emotion} employed a multi-step prompting technique with few training samples for text emotion recognition. \cite{santoso2024large} and \cite{latif2023can} incorporated textual acoustic feature descriptors into prompts. \cite{zhang2024refashioning} investigated several approaches,including in-context learning, few-shot learning, accuracy, generalization, and explanation. \cite{amin2024prompt} examined various prompting techniques, including chain-of-thought, role-play, and their variations.

Despite these advancements, we argue that each approach has its limitations, leaving several concerns unaddressed. For example, \cite{zhang2024refashioning} did not study prompting, \cite{hama2024emotion} only proposed a multi-step prompting without further exploration. The prompting techniques tested by \cite{amin2024prompt} were not specifically designed for emotion. \cite{santoso2024large} and \cite{latif2023can} only considered basic acoustic features, without incorporating more emotion-related properties. Furthermore, both \cite{feng2024foundation} and \cite{gong2023lanser} used ASR transcriptions generated by \textit{Whisper}, whose output has been shown to be robust in emotion recognition even with ASR errors \cite{li2024speech}. However, this setting is not ideal for LLMs handling challenging ASR transcriptions in real-world emotion applications.

\section{Methodology}
To address the issues above, we \textbf{\textit{1)}} Develop prompts that incorporate emotion-specific knowledge; \textbf{\textit{2)}} Incorporate ASR Error Correction (AEC) to refine transcriptions for robust emotion recognition; and \textbf{\textit{3)}} Explore LLM training schemes to further improve the performance.

\subsection{Prompting with Emotion-Specific Knowledge}
In light of the relationship between emotion and relevant disciplines, we extract useful knowledge from acoustics, linguistics, and psychology, to develop emotion-specific prompts for emotion recognition. The prompts are presented in Fig.~\ref{fig:prompt}.

\textbf{Acoustic information} plays a crucial role in distinguishing speech emotions. Features like energy, pitch, and speaking rate have proven useful when been incorporated into prompts as textual descriptors \cite{santoso2024large}. Similarly, gender information, which is highly correlated with pitch, has also been shown to be useful \cite{latif2023can,li2019improved}. However, these features are insufficient to fully describe fine-grained differences in emotions beyond the Big Four. Hence, \textit{we hypothesize that additional acoustic features can enhance LLMs' emotion recognition ability. We propose including pitch range, jitter, and shimmer to incorporate mid-level prosody (between frame-level and utterance-level) and voice quality}.

\begin{figure}
    \centering
    \includegraphics[width=\columnwidth]{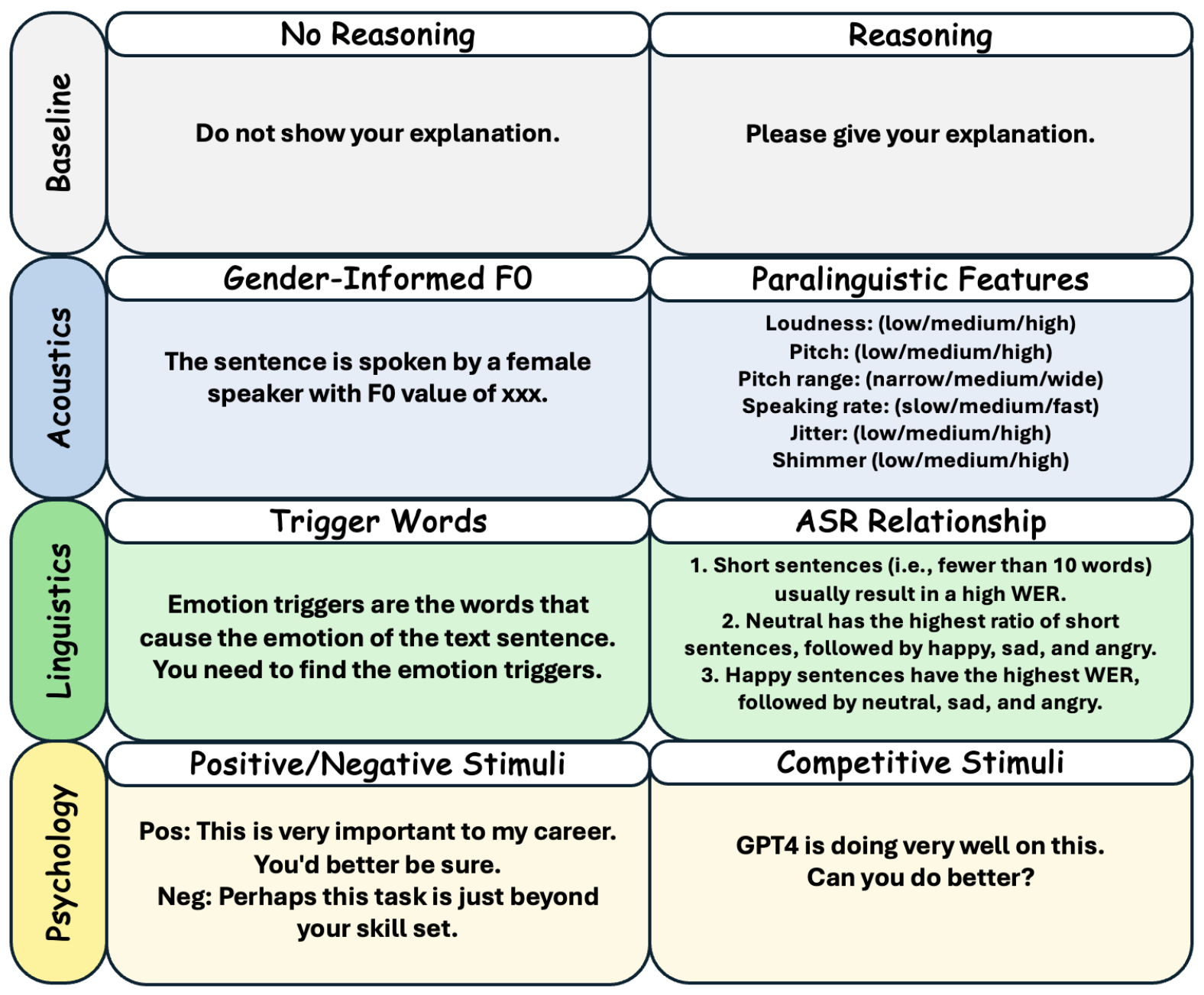}
    \caption{Emotion-specific prompts used in this work.}
    \label{fig:prompt}
\end{figure}

\textbf{Linguistic structure} is essential for understanding emotion in text. For example, \cite{li2023asr} investigated the impact of part-of-speech, affective score, and utterance length on emotion. However, to our knowledge, only one work \cite{singh2023language} has utilized linguistics via identifying emotion triggers (i.e., words that elicit the emotion) when prompting LLMs for emotion prediction. Thus, \textit{we hypothesize that LLM-based emotion recognition can benefit from more linguistic knowledge as LLM processing is inherently text-based. We propose including ASR-emotion relationships among emotion category, Word Error Rate (WER), and utterance length} as outlined in \cite{li2023asr}.

\textbf{Psychological theories}, such as self-monitoring, social cognitive theory, and cognitive emotion regulation have proven effective in improved LLMs' performance across various tasks \cite{li2023large,wang2024negativeprompt}. Therefore, \textit{we hypothesize that LLMs' emotion recognition ability can resonate with their emotional intelligence and thus be enhanced. We propose incorporating positive and negative stimuli from \cite{li2023large,wang2024negativeprompt}, as well as create our novel competitive stimuli.}

\subsection{Emotion Recognition with ASR Error Correction}
Traditional emotion recognition often struggles with imperfect text \cite{li2024speech}. \textit{We argue that it is more challenging to prompt LLMs for emotion recognition on ASR transcription compared to human transcription, due to the presence of word errors}. Therefore, we propose the \textsc{R3} prompting pipeline to perform emotion recognition with AEC and reasoning on ASR transcriptions. The R3 pipeline involves three steps: \textbf{Revise}, where ASR errors are corrected based on N-best hypotheses; \textbf{Reason}, where the LLMs self-explain based on the corrected transcriptions and emotion-specific knowledge; and \textbf{Recognize}, where the emotion is recognized. To incorporate AEC into our prompts, we follow an AEC-specific Alpaca prompt \cite{yang2023generative}, which uses the \textit{``You are an ASR error corrector''} instruction, guiding the LLMs to perform error correction. As LLMs have proven their ability in both AEC and emotion recognition \cite{yang2024large}, this format is expected to facilitate seamless integration with our emotion prompting, instructing the LLMs to function simultaneously as both an ASR error corrector and an emotion recognizer.

\subsection{Exploring LLM Training Schemes}
To understand how LLM training schemes contribute to emotion recognition, we explore context-aware learning, in-context learning, and instruction tuning. For \textbf{context-aware learning}, we organize the sentences in the conversation order and compare different context windows (i.e., the number of sentences preceding the sentence to be recognized). For \textbf{in-context learning}, we test and compare several few-shot cases. For \textbf{instruction tuning}, we apply Parameter-Efficient Fine-Tuning (PEFT) using LoRA \cite{hulora}. We set learning rate, weight decay, and epoch to 1e-4, 1e-5, and 5, respectively, and remain LoRA configuration at its default settings (\href{https://github.com/yc-li20/Emotion-Prompt}{code available}).

\section{Experiments}

\subsection{Datasets and Models}
For the \textbf{datasets}, we use IEMOCAP \cite{busso2008iemocap} and the Test1 set of MSP-Podcast \cite{lotfian2017building}. We combine \textit{excited} with \textit{happy} and use the Big Four classes ({\textit{angry}, \textit{happy}, \textit{neutral}, \textit{sad}}) for IEMOCAP. For MSP-Podcast, we remove \textit{other} and use the eight classes (\textit{angry}, \textit{happy}, \textit{neutral}, \textit{sad}, \textit{disgusted}, \textit{surprised}, \textit{fearful}, \textit{contemptuous}). For the \textbf{ASR models}, we adopt ten popular ones from \cite{li2024speech} to generate diverse transcripts to form 10-best ASR hypotheses. For the \textbf{LLMs}, we utilize \textit{Llama-2 \small{(7b-chat-hf \& 13b-chat-hf)}} and \textit{Falcon \small{(7b-instruct)}}. Temperature and max token are set as 1e-4 and 100. Due to limited space, we mainly present the results on IEMOCAP using \textit{Llama-2} and omit the full prompting message, presenting only the core text following the literature in Sec.~\ref{sec:relatedwork}. Results of using the other model and dataset are presented in necessary experiments.

\subsection{Results and Discussions}
We present the results and discussions based on the following exploration tasks. We replace responses that fall outside the emotion classes with \textit{neutral}. Unweighted Accuracy (UA) is used to measure the results.

\vspace{3pt}
\textit{1. Do WERs have an impact on LLM prompting?}
\vspace{2pt}

In this task, we use the \textbf{baseline no reasoning} prompt: \textit{Predict the emotion from \{the emotion classes\}. Do not show your explanation.} From Table \ref{tab:wer}, we see that WERs do impact LLM prompting. Even the best-performing ASR transcription (i.e., from \textit{Whisper large}) shows more than a 4\% loss compared to ground-truth text. This finding contradicts a previous claim that LLM-based emotion recognition is robust to ASR errors \cite{feng2024foundation}. We believe this discrepancy arises from their (i) use of \textit{Whisper large}, which provides relatively accurate transcriptions, and (ii) introduction of a fifth emotion class, `other', to filter out unconfident labels. However, our setup is more inline with real-world scenarios where emotion recognition is more challenging due to various speaking styles and unconfident labels cannot be filtered. Furthermore, LLM-based performance remains relatively stable within certain WER ranges. The accuracy decrease does not linearly correlate with the WER increase, as seen in traditional deep learning model-based emotion recognition \cite{li2023asr}. Finally, LLMs benefit from more parameters as the 13b model consistently outperforms the 7b model.

\begin{table}[th]
\centering
\large
\caption{Emotion recognition accuracy on transcriptions of increasing WER. $\uparrow$: higher the better. $\downarrow$: lower the better.}
\scalebox{0.72}{
\begin{tabular}{lcc}
\hline
\multirow{2}{*}{\textbf{WER\%$\downarrow$ (\textit{Transcription source})}} & \multicolumn{2}{c}{\textbf{UA\%$\uparrow$}} \\
 & \textbf{\textit{7b-chat-hf}} & \textbf{\textit{13b-chat-hf}} \\ \hline
0.00 (\textit{Ground-truth}) & 44.50 & 47.43 \\ \hdashline
12.3 (\textit{Whisper large}) & 41.77 & 44.27 \\
14.4 (\textit{Whisper small}) & 41.47 & 43.98 \\
20.2 (\textit{Whisper base}) & 41.16 & 43.70 \\
21.9 (\textit{W2V960 large self}) & 41.12 & 43.59 \\
23.8 (\textit{HuBERT large}) & 41.36 & 43.88 \\
26.9 (\textit{Whisper tiny}) & 40.80 & 43.14 \\
27.9 (\textit{W2V960 large}) & 40.49 & 43.10 \\
32.3 (\textit{W2V960}) & 40.00 & 43.01 \\
39.1 (\textit{Wavlm plus}) & 38.01 & 40.12 \\
40.3 (\textit{W2V100}) & 38.09 & 40.19 \\ \hline
\end{tabular}}
\label{tab:wer}
\begin{minipage}{8.5cm}
\centering
\footnotesize \textit{Data: IEMOCAP. LLM: Llama-2.}
\end{minipage}
\end{table}

\vspace{3pt}
\textit{2. Does emotion-specific knowledge help?}
\vspace{2pt}

\begin{table}[ht]
\centering
\caption{Emotion recognition accuracy by using emotion-specific prompts. $\uparrow$: higher the better.}
\begin{tabular}{llcc}
\hline
\multicolumn{2}{c}{\multirow{2}{*}{\textbf{Prompt}}} & \multicolumn{2}{c}{\textbf{UA\%$\uparrow$}} \\
\multicolumn{2}{c}{} & \multicolumn{1}{l}{\textbf{\textit{\small{Ground-truth}}}} & \multicolumn{1}{l}{\textbf{\textit{\small{HuBERT large}}}} \\ \hline
Baseline & 1) No reasoning & 44.50 & 41.36 \\
 & 2) Reasoning & 43.83 \small{($-0.70$)} & 40.07 \small{($-1.29$)} \\
Acoustics & 3) Gender & 45.59 \small{($+1.09$)} & 42.22 \small{($+0.86$)} \\
 & 4) Paraling & 46.25 \small{($+1.75$)} & 43.02 \small{($+1.66$)} \\
Linguistics & 5) Trigger & 46.80 \small{($+2.30$)} & 43.45 \small{($+2.09$)} \\
 & 6) ASR relation & / & 44.10 \small{($+2.74$)} \\
Psychology & 7) Pos stimuli & 45.90 \small{($+1.40$)} & 42.35 \small{($+0.99$)} \\
 & 8) Neg stimuli & 47.43 \small{($+2.93$)} & 42.76 \small{($+1.40$)} \\
 & 9) Cpt stimuli & 45.81 \small{($+1.31$)} & 41.98 \small{($+0.65$)} \\ \hdashline
\multicolumn{2}{c}{Majority voting} & 45.72 \small{($+1.22$)} & 42.94 \small{($+1.58$)} \\
\multicolumn{2}{c}{4 + 5 + 8} & \textbf{48.96 \small{($+4.46$)}} & 44.30 \small{($+2.94$)} \\
\multicolumn{2}{c}{4 + 5 + 6 + 8} & / & \textbf{44.47 \small{($+3.11$)}} \\ \hline
\end{tabular}
\label{tab:prompt}
\begin{minipage}{8.5cm}
\centering
\footnotesize \textit{Data: IEMOCAP. LLM: Llama-2-7b-chat-hf.}
\end{minipage}
\end{table}

\begin{table}[ht]
\centering
\caption{Accuracy comparison with and without Pitch Range (PR), Jitter (Ji), and Shimmer (Sh). $\uparrow$: higher the better.}
\begin{tabular}{lcc}
\hline
\multirow{2}{*}{\textbf{Prompt}} & \multicolumn{2}{c}{\textbf{UA\%$\uparrow$}} \\
 & \textit{\textbf{IEMOCAP}} & \textit{\textbf{MSP-Podcast}} \\ \hline
Baseline no explanation & 44.50 & 35.70 \\
Para info & \textbf{46.25} ($+1.75$) & \textbf{37.37} ($+1.67$) \\
\ \ \ -- w/o Pr, Ji, Sh & 45.89 ($+1.39$) & 36.73 ($+1.03$) \\ \hline
\end{tabular}
\label{tab:para}
\begin{minipage}{8.5cm}
\centering
\footnotesize \textit{LLM: Llama-2-7b-chat-hf.}
\end{minipage}
\end{table}

In this task, we use each of the \textbf{emotion-specific} prompts and their combinations for emotion recognition and compare their effectiveness on both ground-truth and ASR transcriptions. For brevity, we use one ASR transcription, whose WER ranked in the middle, as the representative (i.e., \textit{HuBERT large}). Results are presented in Table.~\ref{tab:prompt}.

We can see that: \textbf{\textit{1)}} All emotion-specific prompts improve the performance, demonstrating the efficacy of our proposed approach by incorporating emotion-specific knowledge. However, the improvement is less pronounced on ASR transcription, highlighting the necessity for AEC. \textbf{\textit{2)}} Our proposed paralinguistic information improves on \cite{santoso2024large}, verifying our hypothesis that additional paralinguistic features are beneficial. Furthermore, the improvement in 8-class is more significant, confirming that these additional features help in distinguishing finer-grained emotions (see Table~\ref{tab:para}). \textbf{\textit{3)}} Linguistic knowledge generally contributes the most, even on ASR transcription. This means that LLMs benefit from identifying emotional trigger words and understanding the ASR-emotion relationship. This ASR-emotion does apply to ground-truth text, as it is specifically developed for ASR transcription. \textbf{\textit{4)}} The steady improvement from psychological knowledge confirms our hypothesis that LLMs' emotion recognition ability can be affected by psychological setting. Interestingly, among the psychological prompts, stimuli with negative affect perform the best. \textbf{\textit{5)}} Surprisingly, the baseline reasoning prompt does not improve performance. By investigating the responses, however, we found this is likely due to the LLM hallucinations, where they often described the (acoustic) tone despite having only text input. \textbf{\textit{6)}} Majority voting underperforms most single prompts, aligning with the finding of \cite{santoso2024large}. Finally, identifying the best prompt combination for both ground-truth and ASR transcriptions, we see that linguistics contributes the most to the latter by having both trigger words and ASR relationships.

\vspace{3pt}
\textit{3. Does the proposed R3 prompt work?}
\vspace{2pt}

In this task, we use the \textbf{R3} prompt: \textit{You are an ASR error corrector and emotion recognizer. Generate the most likely transcript from \{the 10-best ASR hypotheses\} and predict the emotion from \{the emotion classes\} with reasoning based on the provided knowledge.} For comparison, we conduct an ablation study, removing AEC or reasoning. We use \textit{4+5+6+8} as the emotion knowledge since it has proven the best.

\begin{table}[ht]
\centering
\large
\caption{Performance comparison. $\uparrow$: higher the better.}
\scalebox{0.78}{
\begin{tabular}{lccc}
\hline
\multirow{2}{*}{\textbf{Prompt}} & \multicolumn{2}{c}{\textbf{UA\%$\uparrow$}} \\
 & \small{\textbf{\textit{Llama-2-7b}}} & \small{\textbf{\textit{Llama-2-13b}}} & \small{\textbf{\textit{Falcon-7b}}} \\ \hline
R3 & \textbf{49.72} & \textbf{52.27} & \textbf{47.24}  \\
\ \ \ -- w/o AEC & 43.29 & 47.20 & 43.00  \\
\ \ \ -- w/o reasoning & 47.48 & 50.01 & 45.49 \\ \hline
\end{tabular}}
\label{tab:R3}
\begin{minipage}{8.5cm}
\centering
\footnotesize \textit{Data: 10-best of IEMOCAP. Models: -chat-hf \& -instruct.}
\end{minipage}
\end{table}

As shown in Table~\ref{tab:R3}, both AEC and reasoning contribute to the effectiveness of our R3 prompt. Moreover, when incorporating our proposed emotion-specific knowledge, reasoning improves the performance, in contrast to the decrease observed when emotion-specific knowledge was not provided (see Table~\ref{tab:prompt}). This suggests that emotion recognition is particularly challenging for LLMs to reason without relevant information. The examples in Fig.~\ref{fig:r3} illustrate how the R3 prompt helps LLMs in reasoning with emotion-specific knowledge, regardless of whether the recognition is correct.

\begin{figure}
    \centering
    \includegraphics[width=\columnwidth]{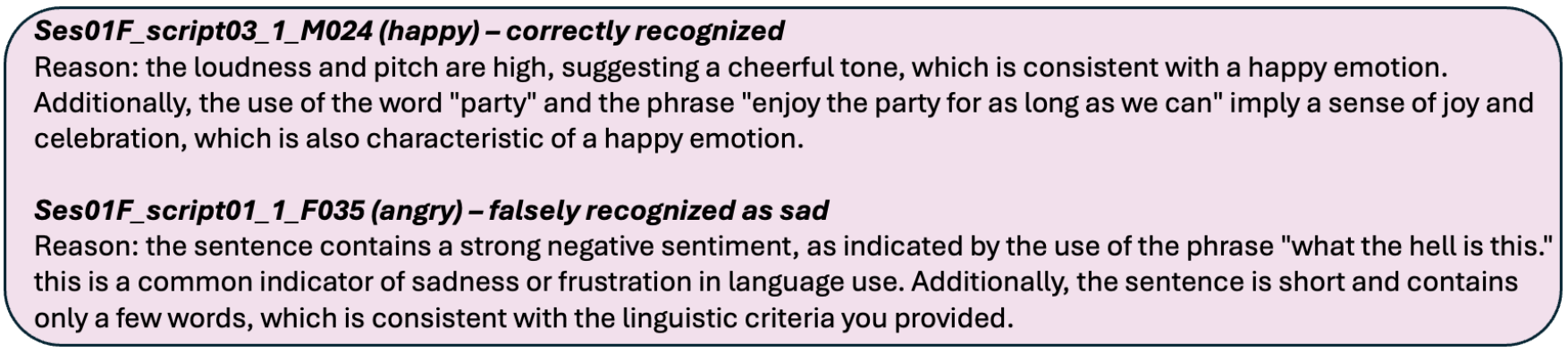}
    \caption{Examples of LLM reasoning with emotion-specific knowledge.}
    \label{fig:r3}
\end{figure}

\vspace{3pt}
\textit{4. Do LLM training schemes help?}
\vspace{2pt}

In this task, we apply the \textbf{R3} prompt with context-aware learning (windows of 5 and 25), in-context learning (5- and 10-shot), and instruction tuning. For instruction tuning, we perform cross-validation by applying PEFT on every four sessions, testing on the remaining session, and then averaging. We do not compare performance across these three approaches due to their different settings.

\begin{table}[ht]
\centering
\large
\caption{Performance via LLM training. $\uparrow$: higher the better.}
\scalebox{0.75}{
\begin{tabular}{lccccc}
\hline
 & \multicolumn{2}{c}{\textbf{\textit{Context-aware}}} & \multicolumn{2}{c}{\textbf{\textit{In-context}}} & {\textbf{\textit{Tuning}}} \\
 & \textbf{\textit{5}} & \textbf{\textit{25}} & \textbf{\textit{5}} & \textbf{\textit{10}} & \textbf{\textit{PEFT}} \\ \hline
\textbf{UA\%$\uparrow$} & 54.35 & 62.46 & 50.74 & 54.36 & 64.67 \\ \hline
\end{tabular}}
\label{tab:training}
\begin{minipage}{8.5cm}
\centering
\footnotesize \textit{Data: 10-best of IEMOCAP. Model: Llama-2-7b-chat-hf.}
\end{minipage}
\vspace{-7pt}
\end{table}

From Table~\ref{tab:training}, it is evident that each LLM training scheme improves the baseline performance of using R3 on 10-best ASR hypotheses (49.72 in Table~\ref{tab:R3}). For context-aware learning and in-context few-shot learning, longer context windows and more samples yield higher accuracy. Instruction tuning leads to the highest performance. Notably, a long context window also results in UA greater than 60\%, indicating the potential to utilize conversational knowledge in real-world LLM-based emotion recognition without tuning the models.

\vspace{3pt}
\textit{5. Are LLMs sensitive to minor prompt variations?}
\vspace{2pt}

In this task, we investigate whether LLM-based emotion recognition is sensitive to minor prompt variations. During our experiments, we observed that prompts with slight differences but the same meaning, such as variations in word choice or the order of provided emotion classes, can largely impact task performance. In Table~\ref{tab:sensitivity}, we modify the \textbf{baseline no reasoning prompt} by changing either the word \textit{Predict} to \textit{Select} or the order of the emotion classes.

\begin{table}[ht]
\centering
\large
\caption{Performance comparison of prompt variations. (A: Angry, H:Happy, N: Neutral, S: Sad). $\uparrow$: higher the better.}
\scalebox{0.75}{
\begin{tabular}{lcccc}
\hline
 & \multicolumn{2}{c}{\textbf{\textit{Word usage}}} & \multicolumn{2}{c}{\textbf{\textit{Emotion order}}} \\
 & \textbf{\textit{Predict}} & \textbf{\textit{Select}} & \textbf{\textit{A, H, N, S}} & \textbf{\textit{H, N, A, S}} \\ \hline
\textbf{UA\%$\uparrow$} & 44.50 & 41.12 & 44.50 & 40.87 \\ \hline
\end{tabular}}
\label{tab:sensitivity}
\begin{minipage}{8.5cm}
\centering
\footnotesize \textit{Data: IEMOCAP. LLM: Llama-2-7b-chat-hf.}
\end{minipage}
\vspace{-2pt}
\end{table}

This aligns with recent findings that LLMs can behave differently due to subtle changes in prompt formatting, such as separators and case, regardless of model size, number of few-shot examples, or instruction tuning \cite{sclarquantifying}. We believe this issue is a major factor hindering the widespread use of LLMs for emotion recognition and similar tasks, thus suggest that future studies evaluating LLMs with prompts would better report the performance across plausible prompt variations.

\section{Conclusion}
In this work, we propose emotion-specific prompts by incorporating relevant knowledge from acoustics, linguistics, and psychology. We also compare LLM-based emotion recognition on both ground-truth and ASR transcriptions, confirming the necessity of AEC. Consequently, we develop the \textsc{Revise-Reason-Recognize} prompting pipeline that integrates AEC, reasoning, and emotion recognition, which proves effective. Additionally, by investigating several LLM training schemes, we confirm the value of longer context windows, more few-shot samples, and instruction tuning. Finally, we uncover the sensitivity of LLMs to minor prompt variations. This research is expected to bridge the gap between existing studies on LLMs and emotion recognition.

\section*{Acknowledgment}

We thank Pinzhen Chen (UoE) for his help with LLM training and Tiantian Feng (USC) for his feedback on LLM selection.

\balance
\bibliographystyle{IEEEbib}
\bibliography{refs}

\end{document}